\documentclass[twocolumn,letterpaper,amsmath,preprintnumbers,amssymb,prb]{revtex4}

\usepackage{graphicx}
\usepackage{dcolumn}
\usepackage{bm}

\begin{document}

\title{Phase diagram of Hertzian spheres}

\author{Josep C. P\`amies$^{a}$}

\author{Angelo Cacciuto$^{a}$}

\author{Daan Frenkel$^{b}$}

\affiliation{$^{a}$Department of Chemistry, Columbia University, 3000 Broadway, New York, NY 10027\\ $^{b}$Department of Chemistry, University of Cambridge, Lensfield Road, Cambridge CB2 1EW, UK and\\ FOM Institute for Atomic and Molecular Physics, Science Park 113, 1098XG Amsterdam, The Netherlands}

\date{\today}

\begin{abstract}
We report the phase diagram of interpenetrating Hertzian spheres. The Hertz potential is purely repulsive, bounded at zero separation, and decreases monotonically as a power law with exponent 5/2, vanishing at the overlapping threshold. This simple functional describes the elastic interaction of weakly deformable bodies and, therefore, it is a reliable physical model of soft macromolecules, like star polymers and globular micelles. Using thermodynamic integration and extensive Monte Carlo simulations, we computed accurate free energies of the fluid phase and a large number of crystal structures. For this, we defined a general primitive unit cell that allows for the simulation of any lattice. We find multiple re-entrant melting and first-order transitions between crystals with cubic, trigonal, tetragonal and hexagonal symmetries.\\
\end{abstract}

\maketitle

\section{Introduction}

Many soft materials consist of building blocks that are themselves ``soft'', meaning that, under experimental conditions, they can interpenetrate or deform appreciably due to the forces acting between them. Linear polymers provide an example:~\cite{soft_colloids} at moderate free-energy cost, the centers of mass of two distinct polymers can be made to superimpose. These macromolecules are therefore fully interpenetrable. Other particles do not interpenetrate but are deformable, like polymer micelles: they deform elastically at high densities, thus exploring configurations that are forbidden to hard-particle systems.

Systems of soft particles are interesting because crystallization in simple and colloidal systems is usually attributed to excluded-volume effects. Systems with soft-repulsive, short-range interactions are therefore expected to have a qualitatively different freezing behavior than those with hard particles.~\cite{Malescio_review, Likos_review} Indeed, one of the striking features of systems of particles with a bounded repulsive interaction is that they have a maximum melting temperature. Above it, the solid phase is unstable. This phenomenon can be understood as follows: consider a system with an interparticle potential that has its maximum energy $\epsilon$ at zero separation and decreases monotonically to zero within a distance $\sigma$. When the thermal energy $k_B\,T\gg \epsilon$, interparticle interactions become negligible and the system approaches ideal gas behavior. However, when $k_B\,T\ll \epsilon$ particles can only overlap at a significant energetic cost. In the limit $k_B\,T/\epsilon\to 0$, the system will undergo a hard-sphere freezing transition. In fact, some 30 years ago, Stillinger already showed that the so-called Gaussian Core Model (GCM) has a maximum melting point.~\cite{Stillinger} Moreover, Stillinger, and subsequently other authors,~\cite{GCM,GCM_accurate} showed that the GCM also exhibits re-entrant melting upon compression, and a transition from face- to body-centered cubic crystal structures. Such a rich phase behavior is not limited to the repulsive Gaussian potential, but has been shown to be common to many effective, soft potentials that were designed to reproduce the phase behavior and dynamics of microgels,~\cite{microgels} star-polyelectrolyte solutions~\cite{PE_star_solutions} and star-polymer solutions.~\cite{star_polymers}

However, not all soft-repulsive potentials give rise to re-entrant melting. The penetrable sphere model (PSM),~\cite{PSM} a square-shoulder potential, is the simplest model that leads to freezing at any arbitrary temperature and to the appearance of cluster crystals (multiple occupation of lattice sites). Multiple occupancy in the PSM arises because, as soon as two particles interpenetrate, no further energetic cost exists for full overlapping. As a consequence, particles can sit on top of each other thus reducing the total number of overlaps and the interaction energy. Density functional theory and molecular simulations have confirmed the occurrence of cluster crystals for the generalized exponential model.~\cite{cluster_crystals} It is expected that certain dendrimers could form such crystals,~\cite{dendrimers} but thus far experiments are lacking (see, however, ref.~\onlinecite{Wigner_crystal_bubble_phases}).

It is apparent from the discussion above that the shape of the pair potential determines when the re-entrant melting and clustering scenarios manifest in the phase diagram. Indeed, Likos {\it et al.}~\cite{Likos_criterion} established a simple criterion to distinguish between the two scenarios for bounded repulsive potentials: re-entrant melting happens for bounded potentials with a positive definite Fourier transform. Otherwise, clustering and freezing are expected to occur at all temperatures.

Although the type of scenario, either re-entrant melting or clustering, can be predicted, there is no method that guarantees an {\it a priori} prediction of all stable crystal structures given the bounded repulsive potential. The usual method involves selecting a small number of candidate structures, for which the ground-state energy or finite-temperature free energies are computed. Among the candidates, the stable structure at any particular density and temperature is the one with the lowest (free) energy. This works well for typical pair potentials, for which one assumes that unusual crystals are unlikely to be stable structures. However, soft potentials are known to induce multiple crystals,~\cite{repulsive_step_potential} including the less typical non-cubic lattices, and exotic structures like the A15 and diamond~\cite{exotic_crystals, crystal_symmetries, geometry_free_energy_soft_spheres} (for an extreme example, see ref.~\onlinecite{2D_repelling_particles}).

There exist powerful approaches to search for stable crystal structures, such as those based on metadynamics~\cite{metadynamics} and genetic algorithms.~\cite{genetic_algorithm} However, metadynamics can be very expensive, in particular when trying to escape deep, local free-energy minima. In a genetic algorithm, an initially random population of lattices is modified using evolutionary rules according to a fitness function: high fitness corresponds to low (free) energy. The structure with the highest fitness in the population after breeding a few generations is assumed to be the most stable lattice. Although genetic algorithms greatly alleviate the problem of searching for candidate crystal structures, they still suffer from the same numerical bottleneck of the usual method: the expensive computation of free energies. For this reason, numerical searches for stable crystal structures are often based on energies, rather than free energies. To improve upon this approach, one can approximate the free energy by that of a harmonic crystal or by using density-functional theory. However, to improve accuracy, one has to turn to computer simulation methods that generate free energies that are ``exact'' to within statistical accuracy. Accurate free-energy calculations can be crucial in soft systems where, specially at high densities, crystals with very similar energies can compete for stability.

In this paper we focus on the phase behavior of Hertzian spheres. The Hertz potential describes the change in elastic energy of two deformable objects when subjected to an axial compression.~\cite{note2} The potential has the following form:
\begin{equation}
    V(r)=\left\{
    \begin{array}{cc}
        \epsilon(1-r/\sigma)^{5/2} & r<\sigma\\
        0 & r\ge\sigma
    \end{array},
    \right.
\end{equation}
where $\sigma$ and $\epsilon$ set the length and energy scales and $r$ is the distance between the centers of the undeformed spheres. In the limit $k_B T/\epsilon\to 0$ the hard-sphere model is recovered. The Hertz model was developed to describe the interaction between elastic spheres that are deformed only slightly. Clearly, for very large compressions or at high densities (well above the overlap concentration) the assumptions underlying the original Hertz model no longer apply. Here we use the Hertz potential as a simple representation of soft particles that is bounded (i.e. it remains finite at $r=0$), has a finite range (viz. $\sigma$) and a positive, yet sharply decaying Fourier transform. Also, the Hertz potential is computationally cheap and thus allows us to perform extensive simulations for the calculation of free energies and coexistence points. We have not attempted to map the present model onto a specific soft-colloid or star-polymer system.~\cite{soft_colloids,star_polymers} We just note that a judicious design of star polymers or dendrimers~\cite{dendrimers} makes it possible to ``engineer'' the interaction between these particles to a considerable degree.

Using standard thermodynamic integration, we calculated free energies of the fluid phase and of candidate crystal structures. As candidates, we considered all Bravais lattices and the hexagonal close-packed, diamond, and A15 structures. Our results show that the phase diagram of the Hertz model exhibits multiple re-entrant melting and polymorphic transitions between a number of crystals. In addition, we find that the re-entrant fluid shows unusual diffusivity curves, and speculate about the nature of this fluid at high density and low temperature.

\begin{figure}
    \includegraphics[height=5.5cm]{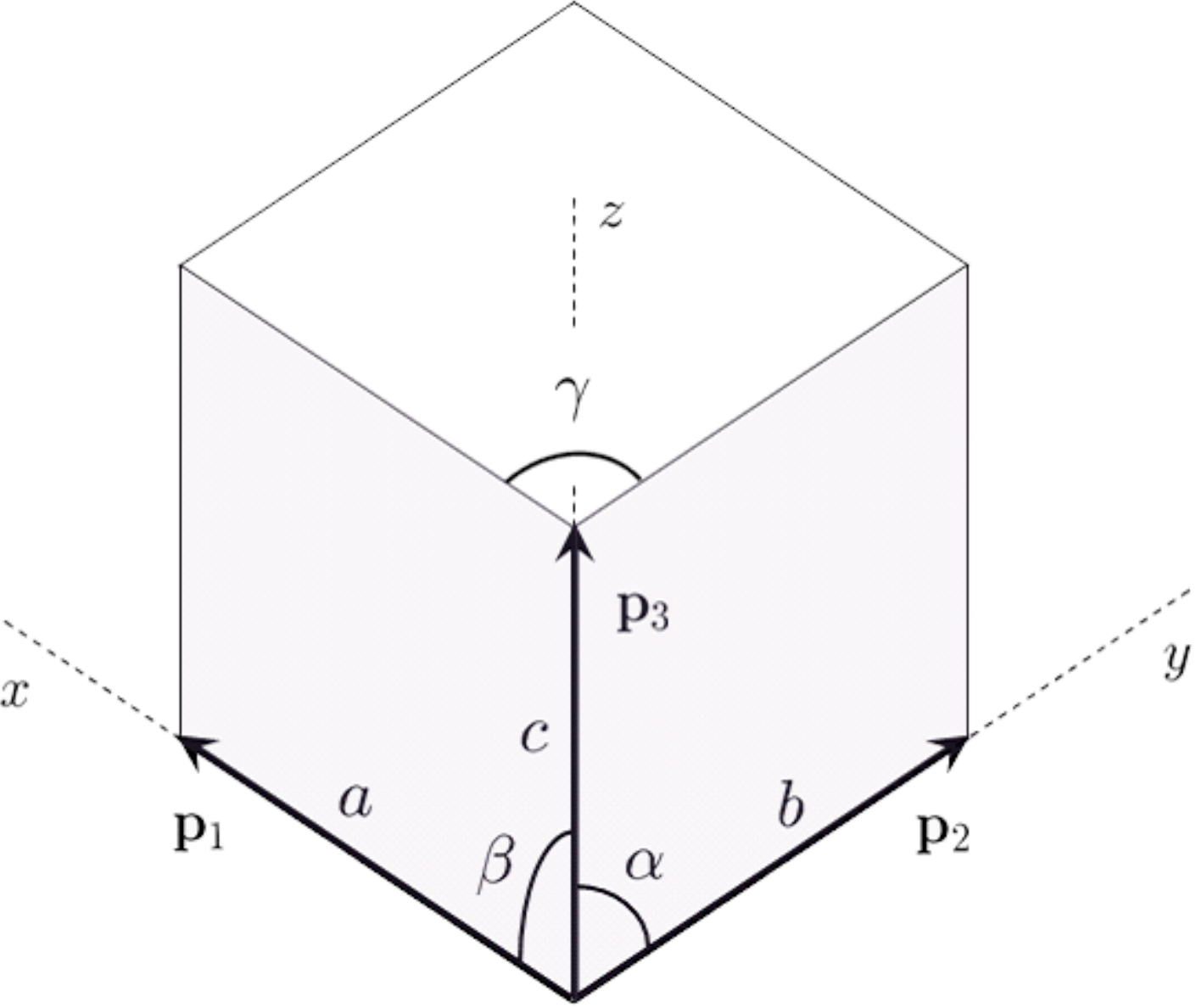}
    \caption{\label{fig:primitive_cell}The general primitive cell and vectors $\mathbf{p}_1$, $\mathbf{p}_2$ and $\mathbf{p}_3$.}
\end{figure}

\begin{table*}
    \caption{\label{table:Bravais_lattices}Parameters characterizing the 14 Bravais lattices according to Eqs.~\ref{eq:position_vector}-\ref{eq:volume}. Round and square brackets indicate free and dependent parameters, respectively.}
	\newcommand\B{\rule{0pt}{2.6ex}}
	\newcommand\A{\rule[-1.2ex]{0pt}{0pt}}
    \begin{ruledtabular}
    \begin{tabular}{l c c c c c c}
        Bravais lattice \A & $b/a$ & $c/a$ & $\alpha/\pi$ & $\beta/\pi$ & $\gamma/\pi$ & $V$ \\
        \hline
        Simple cubic (SC) \B & 1 & 1 & 1/2 & 1/2 & 1/2 & 1 \\
        Face-centered cubic (FCC) & $\sqrt{2}/2$ & $\sqrt{2}/2$ & 1/3 & 1/4 & 1/4 & 1/4\\
        Body-centered cubic (BCC) & 1 & $\sqrt{3}/2$ & 0.304 & 0.304 & 1/2 & 1/2 \\
        Hexagonal (H) & 1 & (1/2) & 1/2 & 1/2 & 1/3 & [$\sqrt{3}/4$] \\
        Simple tetragonal (ST) & 1 & 1/2 & 1/2 & 1/2 & 1/2 & 1/2\\
        Body-centered tetragonal (BCT) & 1 & (3/4) & [0.268] & [0.268] & 1/2 & [1/4]\\
        Rhombohedral or trigonal (R) & 1 & 1 & (1/3) & [1/3] & [1/3] & [$\sqrt{2}/2$]\\
        Simple orthorhombic (SO) & ($\sqrt{2}/2$) & ($\sqrt{3}/2$) & 1/2 & 1/2 & 1/2 & [0.612]\\
        Base-centered orthorhombic (BaCO) & (1) & ($\sqrt{2}/2$) & 1/2 & 1/2 & [1/4] & [1/2]\\
        Body-centered orthorhombic (BCO) & ($\sqrt{2}/2$) & (4/5) & [0.285] & [0.285] & 1/2 & [0.265]\\
        Face-centered orthorhombic (FCO) & (3/5) & ($\sqrt{3}/2$) & [0.340] & [0.304] & [0.186] & [0.235]\\
        Simple monoclinic (SM) & ($\sqrt{2}/2$) & ($\sqrt{3}/2$) & (1/3) & 1/2 & 1/2 & [0.530]\\
        Base-centered monoclinic (BaCM) & 1 & ($\sqrt{2}/2$) & (1/3) & 1/3 & (1/4) & [0.420]\\
        Triclinic (T) & ($\sqrt{2}/2$) & ($\sqrt{3}/2$) & (1/3) & (0.4) & (2/3) & [0.306]\\
    \end{tabular}
    \end{ruledtabular}
\end{table*}

\begin{table*}
    \caption{\label{table:non-Bravais_lattices}Basis vectors for the non-Bravais lattices considered in this work.}
	\newcommand\B{\rule{0pt}{2.6ex}}
	\newcommand\A{\rule[-1.2ex]{0pt}{0pt}}
    \begin{ruledtabular}
    \begin{tabular}{l c c c}
    Non-Bravais lattice \A & Reference Bravais lattice & $n_b$ & $\mathbf{b}_i/a$ \\
    \hline
    \vspace{0.15cm}
    Diamond (D) \B & FCC & 1 & $\left(\frac{1}{4},\frac{1}{2},\frac{1}{2}\right)$ \\
    \vspace{0.15cm}
    Hexagonal close-packed (HCP) & Simple hexagonal (SH) (H with $c/a=1$) & 1 & $\left(\frac{1}{2},\frac{1}{3},\frac{1}{2}\right)$ \\
    \vspace{0.15cm}
    A15 & SC & 7 & $(\frac{1}{2},\frac{1}{2},\frac{1}{2}) \left(\frac{1}{2},0,\frac{1}{4}\right) \left(\frac{1}{2},0,\frac{3}{4}\right) \left(\frac{1}{4},\frac{1}{2},0\right)$ \\
    & \A & & $\left(\frac{3}{4},\frac{1}{2},0\right) \left(0,\frac{1}{4},\frac{1}{2}\right) \left(0,\frac{3}{4},\frac{1}{2}\right)$ \\
    \end{tabular}
    \end{ruledtabular}

\end{table*}

\section{General primitive unit cell}

\subsection{General framework for Bravais lattices}
\label{sec:Bravais}

All 14 Bravais lattices can be generated from a general primitive unit cell~\cite{note1} defined by three independent vectors $\mathbf{p}_1$, $\mathbf{p}_2$ and $\mathbf{p}_3 \in \mathbb{R}^3$. Once these primitive vectors are defined for a particular crystal, the location of every particle in the crystal follows from
\begin{equation}
    \mathbf{r} = n_1\mathbf{p}_1+n_2\mathbf{p}_2+n_3\mathbf{p}_3+\mathbf{r}_0,
    \label{eq:position_vector}
\end{equation}
where $n_1, n_2$ and $n_3 \in \mathbb{Z}$, and $\mathbf{r}_0$ is an arbitrary vector. For simplicity, we can fix one of the particles at the origin of the coordinate system, and thus $\mathbf{r}_0 = (0,0,0)$.

There are many ways of defining primitive vectors for each Bravais lattice. Here we conveniently choose the general set of vectors depicted in Fig.~\ref{fig:primitive_cell}, where $\mathbf{p}_1$ is parallel to the $x$ axis, and $\mathbf{p}_1$ and $\mathbf{p}_2$ lie on the $xy$ plane of the cartesian coordinate system. Algebraically,
\begin{eqnarray}
    \label{eq:primitive_vectors1}
    \frac{1}{a}\,\mathbf{p}_1 & = & (1,0,0),\\
    \frac{1}{a}\,\mathbf{p}_2 & = & \frac{b}{a}(\cos\gamma,\sin\gamma,0),\\
    \frac{1}{a}\,\mathbf{p}_3 & = &
    \label{eq:primitive_vectors3}
    \frac{c}{a}(\cos\beta,\cos\alpha'\sin\beta,\sin\alpha'\sin\beta),
\end{eqnarray}
where $a$, $b$, and $c$ are the lengths of $\mathbf{p}_1$, $\mathbf{p}_2$ and $\mathbf{p}_3$,  respectively, and $\alpha$, $\beta$ and $\gamma$ denote the angles formed by each pair of primitive vectors, as indicated in Fig.~\ref{fig:primitive_cell}. The angle $\alpha'$ is such as
\begin{eqnarray}
    &&\cos \alpha' = \frac{\cos\alpha - \cos\beta\cos\gamma}{\sin\beta\sin\gamma},\\
    &&\sin\alpha' = \nonumber\\
    & = & \frac{\sqrt{-\cos^2\alpha-\cos^2\beta+\sin^2\gamma+2\cos\alpha\cos\beta\cos\gamma}}{\sin\beta\sin\gamma}.\;\;\;\;\;\;
    \label{eq:alpha}
\end{eqnarray}

The volume $V$ of the primitive cell is $|\mathbf{p}_1\cdot(\mathbf{p}_2\times\mathbf{p}_3)|$ and, therefore,
\begin{equation}
    \frac{V}{a^3} = \frac{b}{a}\frac{c}{a}\sin\alpha'\sin\beta\sin\gamma.
    \label{eq:volume}
\end{equation}

\noindent
From eq.~\ref{eq:volume} we can see that the length $a$ sets the specific volume of the crystal. Without loss of generality, we choose $a$ to be the longest among the three primitive vectors. Hence, according to eqns.~\ref{eq:position_vector}-\ref{eq:alpha}, the 14 Bravais lattices can be described with five parameters: $0<b/a\leq1$, $0<c/a\leq1$, $0<\alpha\leq\pi/2$, $0<\beta\leq\pi/2$ and $0<\gamma\leq\pi$. Table~\ref{table:Bravais_lattices} shows the specific values of the parameters for the 14 Bravais lattices. Depending on the particular lattice, any of the five parameters can be free, fixed or dependent. A free parameter, shown in round brackets in Table~\ref{table:Bravais_lattices}, can be changed at will to generate different specific structures belonging to the same lattice, provided that $-\cos^2\alpha-\cos^2\beta+\sin^2\gamma+2\cos\alpha\cos\beta\cos\gamma\geq0$. A dependent parameter, shown in square brackets, has values that depend on those assigned to the free parameters, according to the expressions in Table~\ref{table:lattice_dependent_parameters}. By definition, some lattices are a subset of the tetragonal, trigonal, orthorhombic, monoclinic or triclinic, as indicated in Table~\ref{table:lattice_equivalents}.

\subsection{Extension to non-Bravais lattices}
\label{sec:non-Bravais}

Bravais lattices have 1 particle in their primitive unit cell. The primitive unit cell of non-Bravais crystals contains $n_b$ additional particles that are characterized by a set of translation or basis vectors $\mathbf{b}_i$, where $i$ runs from $0$ to $n_b$. Because of translational symmetry, one of the basis vectors can be chosen arbitrarily, and this is why in eq.~\ref{eq:position_vector} we defined $\mathbf{b}_0=\mathbf{r}_0=(0,0,0)$. A non-Bravais crystal is generated by translation of its reference Bravais lattice using the set of basis vectors
\begin{eqnarray}
    \left\{\mathbf{r}\right\}^{\text{NB}} & = & \sum_{i=0}^{n_b}
    (\left\{\mathbf{r}\right\}^{\text{B}} + \mathbf{b}_i).
\end{eqnarray}
Table~\ref{table:non-Bravais_lattices} shows the reference lattice and basis vectors corresponding to the three non-Bravais crystals considered in this work.

\subsection{General periodic boundary conditions}

According to the framework described in sections \ref{sec:Bravais} and \ref{sec:non-Bravais}, one can generate any crystal structure as a periodic repetition of the primitive unit cell along its three primitive vectors. Equivalently, one can also construct a unit cell that consists of an arbitrary number of primitive cells, and replicate the unit cell along its lattice vectors to obtain the same periodic structure. Therefore, we can define a non-primitive unit cell that contains a sufficiently large number of particles, and use general periodic boundary conditions to simulate the bulk crystal. The lattice vectors that define this unit cell are $\mathbf{u}_1 = n_1 \mathbf{p}_1$, $\mathbf{u}_2 = n_2 \mathbf{p}_2$ and $\mathbf{u}_3 = n_3 \mathbf{p}_3$, where $n_1$, $n_2$ and $n_3$ are the number of primitive cells contained in the unit cell in each of the directions of the lattice vectors.

The nearest image of a particle in a unit cell can be found using the reciprocal lattice vectors $\mathbf{v}_1$, $\mathbf{v}_2$ and $\mathbf{v}_3$, which are defined such as $\mathbf{u}_i \cdot \mathbf{v}_j=\delta_{ij}$. Using the general primitive vectors of eqns.~\ref{eq:primitive_vectors1}-\ref{eq:primitive_vectors3}, the corresponding reciprocal vectors can be written as
\begin{eqnarray}
    a\,\mathbf{v}_1 & = & \frac{\mathbf{u}_2 \times \mathbf{u}_3}
    {(\mathbf{u}_2 \times \mathbf{u}_3) \cdot \mathbf{u}_1} =  \nonumber\\
    & = & \left(1,\frac{-\cos\gamma}{\sin\gamma},\frac{\cos\alpha'\sin\beta\cos\gamma-\cos\beta\sin\gamma}{\sin\alpha'\sin\beta\sin\gamma}\right),\\
    a\,\mathbf{v}_2 & = & \frac{\mathbf{u}_3 \times \mathbf{u}_1}
    {(\mathbf{u}_3 \times \mathbf{u}_1) \cdot \mathbf{u}_2} =
  \frac{1}{b/a}\left(0,\frac{1}{\sin\gamma},\frac{-\cos\alpha'}{\sin\alpha'\sin\gamma}\right),\;\;\;\;\;\;\\
    a\,\mathbf{v}_3 & = & \frac{\mathbf{u}_1 \times \mathbf{u}_2}
    {(\mathbf{u}_1 \times \mathbf{u}_2) \cdot \mathbf{u}_3} =
    \frac{1}{c/a}\left(0,0,\frac{1}{\sin\alpha'\sin\beta}\right).
\end{eqnarray}

Following eq.~\ref{eq:position_vector}, the location $\mathbf{r}_u$ of a particle inside the unit cell with an image at position $\mathbf{r}$ can be calculated as
\begin{eqnarray}
    \mathbf{r}_u & = &
    \mathbf{r}-m_1\mathbf{u}_1-m_2\mathbf{u}_2-m_3\mathbf{u}_3,
\end{eqnarray}
where
\begin{eqnarray}
    m_i=\left\{\mathbf{r}\cdot\mathbf{v}_i\right\}\;\text{for}\;
    i=1,2,3
\end{eqnarray}
and $\left\{\,\cdots\right\}$ denotes the nearest integer. Note that, by definition, $\left\{\mathbf{r}_u\cdot\mathbf{v}_i\right\}=0$.

\begin{table}
    \caption{\label{table:lattice_dependent_parameters}Lattice-dependent parameters.}
	\newcommand\B{\rule{0pt}{3.2ex}}
	\newcommand\A{\rule[-1.2ex]{0pt}{0pt}}
    \begin{ruledtabular}
    \begin{tabular}{l c}
    Bravais lattice \A & Dependent parameters \\
    \hline
    BCT, BCO \B & $\alpha=\beta=\arccos\left[\frac{1}{2\frac{c}{a}}\right]$ \\
    \vspace{0.25cm}
    R \B & $\beta=\gamma=\alpha$ \\
    \vspace{0.25cm}
    FCO & $\alpha=\arccos\left[\frac{\left(\frac{b}{a}\right)^2+\left(\frac{c}{a}\right)^2-\left(\frac{b}{a}\sin\gamma\right)^2-\left(\frac{c}{a}\sin\beta\right)^2}{2\frac{b}{a}\frac{c}{a}}\right]$ \\
    \vspace{0.25cm}
    FCO & $\beta=\arccos\left[\frac{1}{2\frac{c}{a}}\right]$ \\
    BaCO, FCO & $\gamma=\arccos\left[\frac{1}{2\frac{b}{a}}\right]$ \\
    \end{tabular}
    \end{ruledtabular}
\end{table}

\begin{table}
    \caption{\label{table:lattice_equivalents}Parameters leading to equivalent lattices.}
	\newcommand\B{\rule{0pt}{2.6ex}}
	\newcommand\A{\rule[-1.2ex]{0pt}{0pt}}
    \begin{ruledtabular}
    \begin{tabular}{l r}
    Bravais lattice \A & Parameters and equivalent lattice \\
    \hline
    ST \B & $c/a=1 \mapsto$ SC \\
    BCT & $c/a=\sqrt{3}/2 \mapsto$ BCC \\
    R & $\alpha=\pi/2 \mapsto$ SC \\
    SO & $b/a=1$ or $c/a=1$ or $b/a=c/a \mapsto$ ST \\
    & $b/a=c/a=1 \mapsto$ SC \\
    BaCO & $b/a=c/a=\sqrt{2}/2 \mapsto$ SC \\
    BCO & $b/a=1 \mapsto$ BCT \\
    & $b/a=1$ and $c/a=\sqrt{3}/2 \mapsto$ BCC \\
    FCO & $b/a=c/a=\sqrt{2}/2 \mapsto$ FCC \\
    SM & $\alpha=\pi/2 \mapsto$ SO \\
    BaCM & $c/a=1$ and $\alpha=\gamma=1/3 \mapsto$ R with $\alpha=\pi/3$ \\
    T & parameters of Table \ref{table:Bravais_lattices} $\mapsto$ any of the lattices \\
    \end{tabular}
    \end{ruledtabular}
\end{table}

\begin{figure*}
    \includegraphics[scale=.475]{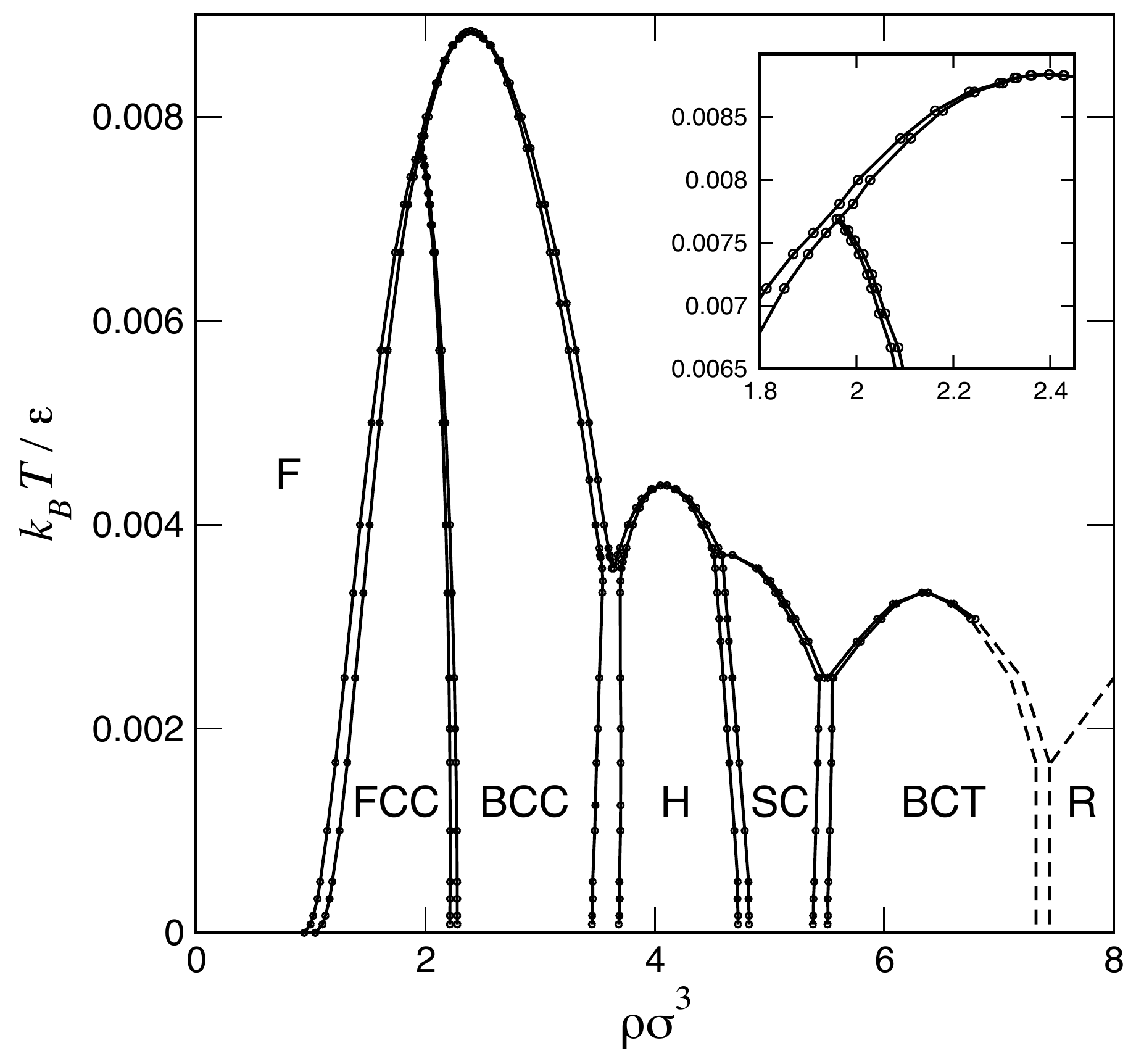}
    \includegraphics[scale=.475]{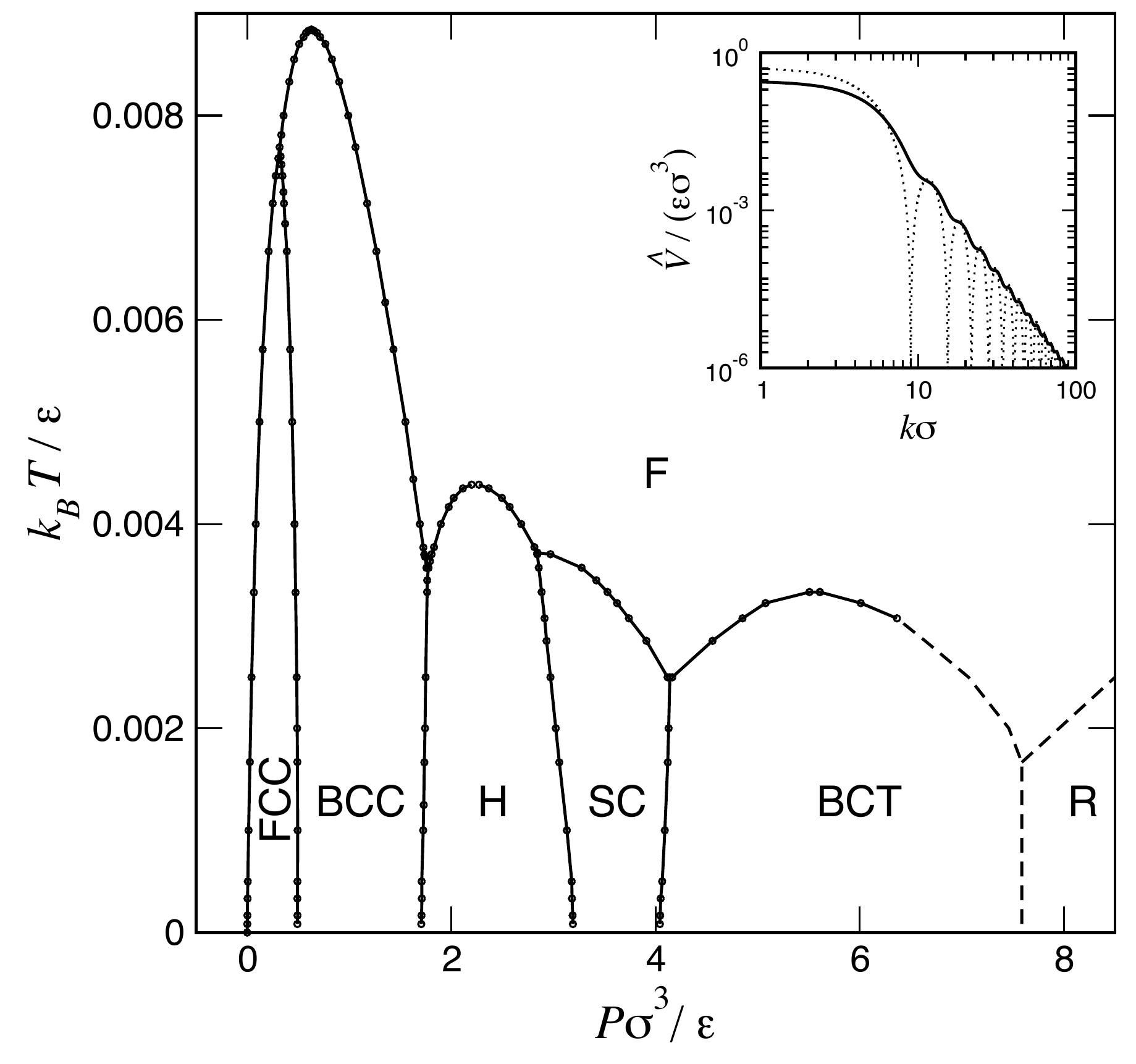}
    \caption{\label{fig:phase_diagram} Temperature-density and temperature-pressure cuts of the phase diagram of Hertzian spheres. Errors are smaller than the size of the circles. Solid lines are a guide to the eye. We performed accurate free-energy calculations for $\rho\sigma^3<7$. Broken lines thus indicate approximate phase boundaries. The inset on the left zooms in the region around the F-FCC-BCC triple point, and the inset on the right compares the Fourier transform of the Hertz (solid line) and 3D overlap (dotted line) potentials. See section~\ref{sec:Discussion} for details.}
\end{figure*}

\begin{table*}
    \caption{\label{table:coexistence_points}Some phase-coexistence points. Standard deviations for densities are below $5\,10^{-3}$.
    }
	\newcommand\B{\rule{0pt}{2.6ex}}
	\newcommand\A{\rule[-1.2ex]{0pt}{0pt}}
    \begin{ruledtabular}
    \begin{tabular}{l c c c c c c c c c c c}
    Phases 1-2 \A & $\epsilon$ & $10^3\;k_BT/\epsilon$ & $P\sigma^3/k_BT$ & $\rho_1\sigma^3$ & $\rho_2\sigma^3$ & \;\;\;\; Phases 1-2 & $\epsilon$ & $10^3\;k_BT/\epsilon$ & $P\sigma^3/k_BT$ & $\rho_1\sigma^3$ & $\rho_2\sigma^3$ \\
    \hline
    F - FCC \B & 12000 & $\;\;\;\;\;\;\;\;$8.333\,$10^{-2}$ & 12.34 & 1.000 & 1.104 & \;\;\;\; F - H & 280 & 3.571 & 497.4 & 3.644 & 3.709 \\
    F - FCC & 1000 & 1.000 & 14.74 & 1.145 & 1.252 & \;\;\;\; F - H & 250 & 4.000 & 474.5 & 3.764 & 3.811 \\
    F - FCC & 400 & 2.500 & 17.78 & 1.295 & 1.389 & \;\;\;\; F - H & 228 & 4.386 & 501.2 & 4.046 & 4.050 \\
    F - FCC & 200 & 5.000 & 24.15 & 1.531 & 1.602 & \;\;\;\; H - F & 270 & 3.704 & 767.4 & 4.579 & 4.518 \\
    F - BCC & 125 & 8.000 & 44.75 & 2.003 & 2.028 & \;\;\;\; H - F & 240 & 4.167 & 617.1 & 4.361 & 4.325 \\
    F - BCC & 113.1 & 8.842 & 71.17 & 2.398 & 2.398 & \;\;\;\; H - SC & 12000 & $\;\;\;\;\;\;\;\;$8.333\,$10^{-2}$ & 38290 & 4.725 & 4.822 \\
    FCC - BCC & 12000 & $\;\;\;\;\;\;\;\;$8.333\,$10^{-2}$ & 5912 & 2.214 & 2.279 & \;\;\;\; H - SC & 1000 & 1.000 & 3132 & 4.693 & 4.785 \\
    FCC - BCC & 1000 & 1.000 & 494.6 & 2.216 & 2.275 & \;\;\;\; H - SC & 400 & 2.500 & 1189 & 4.597 & 4.676 \\
    FCC - BCC & 400 & 2.500 & 194.3 & 2.206 & 2.252 & \;\;\;\; H - SC & 270 & 3.704 & 768.6 & 4.518 & 4.586 \\
    FCC - BCC & 200 & 5.000 & 88.35 & 2.148 & 2.175 & \;\;\;\; SC - F & 270 & 3.704 & 802.4 & 4.674 & 4.677 \\
    FCC - BCC & 150 & 6.667 & 58.23 & 2.071 & 2.086 & \;\;\;\; SC - F & 325 & 3.077 & 1215 & 5.184 & 5.227 \\
    FCC - BCC & 130 & 7.692 & 41.28 & 1.959 & 1.966 & \;\;\;\; SC - F & 400 & 2.500 & 1647 & 5.419 & 5.478 \\
    BCC - F & 280 & 3.571 & 491.9 & 3.540 & 3.625 & \;\;\;\; SC - BCT & 12000 & $\;\;\;\;\;\;\;\;$8.333\,$10^{-2}$ & 48510 & 5.379 & 5.506 \\
    BCC - F & 200 & 5.000 & 310.1 & 3.355 & 3.427 & \;\;\;\; SC - BCT & 1000 & 1.000 & 4091 & 5.402 & 5.526 \\
    BCC - F & 140 & 7.143 & 164.6 & 2.997 & 3.045 & \;\;\;\; F - BCT & 400 & 2.500 & 1665 & 5.508 & 5.558 \\ 
    BCC - F & 125 & 8.000 & 123.9 & 2.808 & 2.840 & \;\;\;\; F - BCT & 350 & 2.857 & 1595 & 5.760 & 5.799 \\
    BCC - H & 12000 & $\;\;\;\;\;\;\;\;$8.333\,$10^{-2}$ & 20470 & 3.453 & 3.686 & \;\;\;\; F - BCT & 310 & 3.226 & 1574 & 6.076 & 6.105 \\
    BCC - H & 1000 & 1.000 & 1724 & 3.477 & 3.701 & \;\;\;\; BCT - F & 325 & 3.077 & 2069 & 6.750 & 6.796 \\
    BCC - H & 400 & 2.500 & 699.5 & 3.516 & 3.699 & \;\;\;\; BCT - F & 300 & 3.333 & 1683 & 6.381 & 6.384 \\
    \end{tabular}
    \end{ruledtabular}
\end{table*}

\section{Thermodynamic integration}

In order to construct the phase diagram, we first performed a rough test of the stability of the fluid phase by running short NVT Monte Carlo simulations at 100 points equally distributed in the $T-\rho$ space, $k_B\;T/\epsilon$ within the interval $[0-0.01]$ and $\rho\sigma^3$ within $[0-10]$. This allowed us to map approximately the solid region of the diagram. Secondly, we distributed 28 points along the isotherm at $\epsilon/k_BT=600$ for $\rho\sigma^3$ within $[1-8]$, and tested the mechanical stability of all Bravais crystals and the three non-Bravais structures considered by running a short NVT simulation for each specific structure. In particular, for lattices with free parameters, typically simulations at 5 different values for each free parameter were carried out. This large set of simulations reduced significantly the number of crystal structures for which we subsequently computed free energies accurately: in particular, at $\rho\sigma^3=4$ we calculated free energies for the SC, BCC, FCC, R and H. At $\rho\sigma^3=5$, for the SC, H, and R. At $\rho\sigma^3=7$, for the ST, BCT, BCO, H, and R.

To determine the thermodynamically-stable crystals, we obtained free energies for the set of mechanically-stable candidate structures. We employed standard thermodynamic integration~\cite{Frenkel-Ladd, Frenkel&Smit} using the Einstein crystal as a reference state and a 20-point Gauss-Legendre numerical integration. To assess accuracy, particularly at high densities, we carried out four integrations for each crystal structure, at values of $\alpha\sigma^2/k_BT$ equal to 50, 100, 150 and 200, where $\alpha$ is the Einstein-crystal spring constant. The calculated free energies are thus averages over four integrations. Our NVT simulations typically consisted of 500-700 particles, the actual number depending on the crystal structure. In particular, we used 500 spheres for the FCC, 686 for the BCC, 512 for the H and SC, and 648 for the BCT. Occasionally, simulations with about 1000 particles were run at a few points distributed in the $T-\rho$ space, but we did not find any significant finite-size effects.

The determination of coexistence points involved NPT simulations to compute the local equation of state $P(\rho)$ around each pair of guesses for the equilibrium densities. We used 30 simulation runs for each guess and fitted a 10-degree polynomial function to the data. The free energy of the solid phase at the estimated density was calculated via integration from the Einstein crystal as explained in the paragraph above. We computed the free energy of the fluid by integrating energies from 800-particle NVT simulations along a constant density path starting at the isotherm at $k_BT/\epsilon=1/105$. This isotherm consisted of polynomial fits to the equation of state of the fluid for eight different equally-spaced intervals of $\rho\sigma^3$ within the range $[0-8]$. We obtained free energies along the isotherm by integrating the equation of state from the ideal-gas limit at zero density. We generated as many simulation points as necessary to keep standard deviations of integrals below 0.01\%. Accuracy was absolutely important also when comparing free energies of competing crystals, since differences were often below $5\,10^{-3} k_BT$ per particle.

Once the local equations of state $P(\rho)$ and chemical potentials $\mu(\rho)$ at fixed $T$ were determined, we searched for the densities $\rho_1$ and $\rho_2$ that satisfy $P_1(\rho_1)=P_2(\rho_2)$ and $\mu_1(\rho_1)=\mu_2(\rho_2)$. We then checked that $\rho_1$ and $\rho_2$ fall within the interval for which $P(\rho)$ and $\mu(\rho)$ were fitted. Otherwise, we used $\rho_1$ and $\rho_2$ as new guesses for the coexistence densities, and repeated the procedure explained in the previous paragraph.

\section{Phase diagram}

According to the criterion established by Likos in ref.~\onlinecite{Likos_criterion}, soft potentials that are purely repulsive and bounded give rise to re-entrant melting when its Fourier transform exists and it does not have negative components. If the Fourier transform oscillates around zero, cluster solids appear instead. The Fourier transform of the Hertz potential is definite positive and, indeed, our results confirm the re-entrant melting scenario, as Fig.~\ref{fig:phase_diagram} shows. Table~\ref{table:coexistence_points} summarizes phase coexistence data.

The phase diagram of Hertzian spheres shows that the fluid phase freezes upon compression to form an FCC crystal which, at higher densities, turns into a BCC structure. However, the BCC packing is favored over the FCC at high temperatures because particles in the former have a higher vibrational entropy. An inset in Fig.~\ref{fig:phase_diagram} shows in detail the area around the F-FCC-BCC triple point, which occurs at $k_BT/\epsilon=7.69\,10^{-3}$. The FCC phase does not have a maximum freezing point for the Hertz potential, as opposed to the GCM.~\cite{GCM_accurate} The fluid re-enters at densities larger than the maximum freezing point at $k_BT/\epsilon=8.84\,10^{-3}$ and $\rho\sigma^3=2.40$. At lower temperatures and with increasing density, we find multiple re-entrant melting and four more crystals: hexagonal, simple cubic, body-centered-tetragonal and trigonal. Simple cubic crystals of one component are rare, because they are usually mechanically unstable.

\begin{figure}
    \includegraphics[scale=.45]{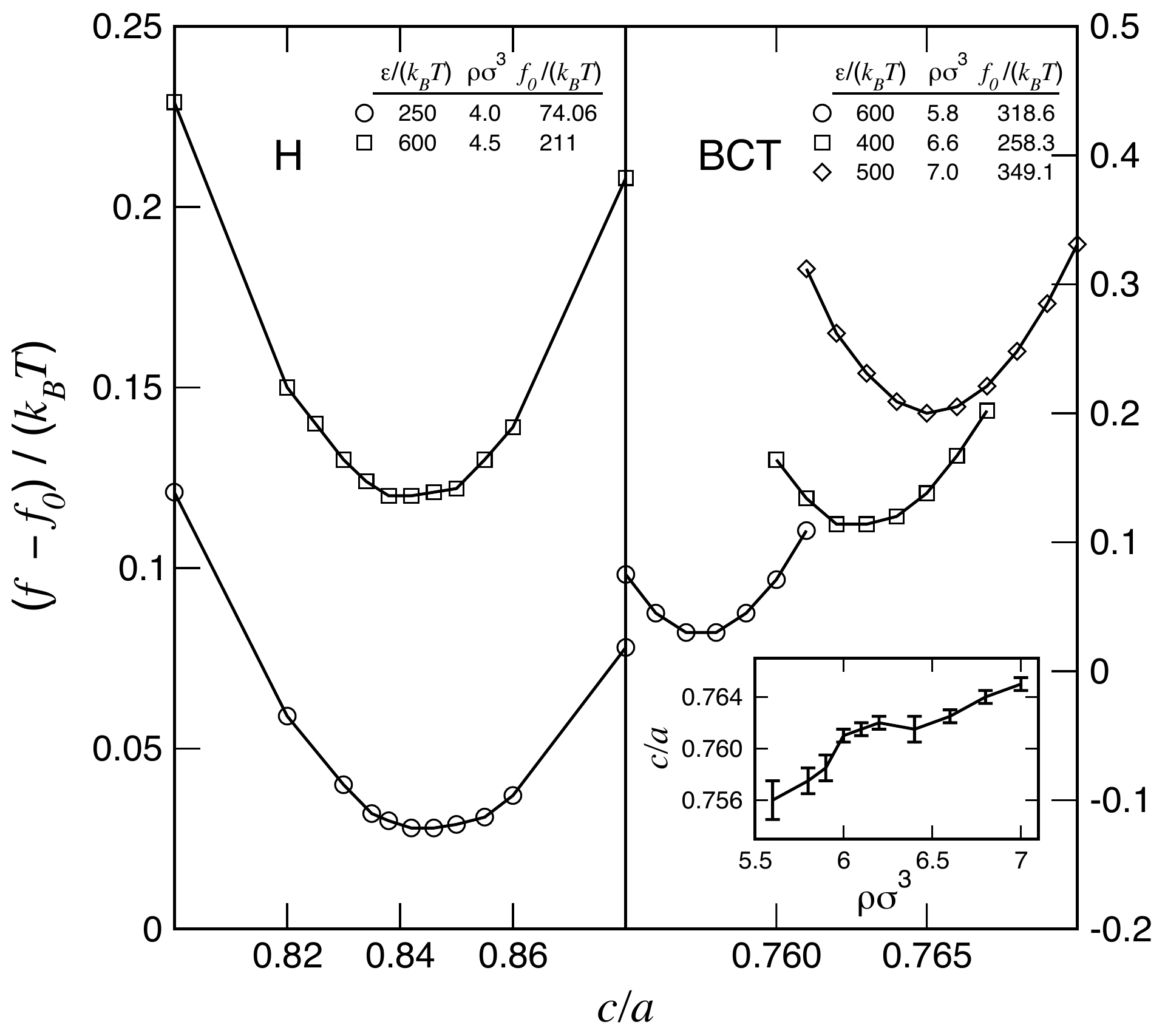}
    \caption{\label{fig:free_energies} Free energies per particle of the hexagonal (left) and body-centered tetragonal (right) crystals, as a function of the respective aspect ratio $c/a$ of the primitive unit cell.  Free-energy sets have been displaced arbitrarily by $f_0$ for each specific point in the phase diagram, as indicated in the legends. Errors are smaller than the size of the symbols. The optimal $c/a$, at which the free energy is at a minimum, is density dependent for the BCT structure. The inset shows that the density behavior of the optimal $c/a$ has inflection points such that aspect ratios close to that of the BCT at the maximum melting point are favored. Lines are a guide to the eye.}
\end{figure}

With regard to the H and BCT structures, the geometry of their primitive cells is not unique, as explained in section \ref{sec:Bravais}. Both crystals have one degree of freedom, namely the aspect ratio of the primitive cell $c/a$ (see Table~\ref{table:Bravais_lattices}). One expects that the {\it optimal} value of $c/a$, at which the free energy reaches a minimum, is in general density dependent. We show this in Fig.~\ref{fig:free_energies}, where we plotted free energies per particle as a function of the aspect ratio. For the H crystal, the minimum in free energy occurs at $c/a \approx 0.84$, independently of temperature. Within the range of densities at which the H structure is stable, we do not observe any significant density dependence either, as the free-energy curve is relatively flat around the minimum. However, this is not the case for the BCT. As shown in the inset of Fig.~\ref{fig:free_energies}, the maximum freezing point, occurring at $\rho\sigma^3=6.35$, leaves its ``imprint'' on the density dependence of the optimal aspect ratio of the BCT primitive cell, as there is a clustering of aspect ratios close to $c/a=0.762$, which is the value at the maximum freezing point.

\section{Discussion}
\label{sec:Discussion}

Very recently, Prestipino {\it et al.}~\cite{Hertz_ground_states} calculated the ground-state structures for the Hertz potential. They found that the non-bravais cI16 and A5 lattices minimize the energy at densities within the interval $[4.3-6.4]$. We did not include the cI16 and A5 in the set of candidate structures, and hence we cannot tell whether the cI16 and A5 remain stable at $T>0$. However, we stress that entropic contributions are not negligible even at the lowest temperatures that we studied. For example, at $\rho\sigma^3=5$, the trigonal crystal with $\alpha/\pi=0.46$ yields the lowest energy of all the candidate structures considered in this work. Yet, the simple cubic structure has the lowest free energy; it is thus stabilized by entropy. Furthermore, as discussed in detail below, the phase diagram and both structural and dynamical anomalies seen in the fluid are interconnected. For example, the maximum melting point of the BCT lies at $\rho\sigma^3=6.35$, which is approximately the density at which the radial distribution function and the diffusion coefficient reach local extrema. But at $T=0$ a transition between the A5 and the BCT happens to be close to that same density~\cite{Hertz_ground_states} (the A5 becomes unstable at $\rho\sigma^3=6.386$), which makes an extremum occurring at the same location in the phase diagram unlikely. The fact that there is such a strong correlation between the shape of the computed phase diagram and the dynamics of the fluid provides indirect support for the assumption that we have indeed identified the stable crystal phases at $T>0$.

In the next paragraphs we discuss a few interesting aspects of the fluid phase. The re-entrance of the fluid at high densities and low temperatures (e.g. at $\rho\sigma^3=7.5$ and $k_BT/\epsilon=2.5\,10^{-3}$; see Fig.~\ref{fig:phase_diagram}) appears to be different in nature from that of a hard-sphere fluid, which has a rugged potential-energy landscape and yet is a fluid because there is an extended and connected region in configuration space where the potential energy is strictly zero (and therefore flat). For the Hertz model at the aforementioned density, the typical inter-particle distance in the liquid phase is of the order of 0.5$\sigma$ and the potential is at less than 20\% of its maximum value; also, at the aforementioned temperature, the relative radial displacement of two particles by as little as $2.5\,10^{-3}\sigma$ will change the potential energy of such a pair by an amount that is comparable to $k_BT$. Hence, the potential-energy landscape is far from being flat at these conditions. We suggest that the reason why Hertzian spheres (and other potentials showing re-entrant melting, like the GCM) melt at high densities and low temperatures has to do with the fact that the Fourier transform of the potential goes to zero rapidly. Indeed, S\"ut\H{o}~\cite{ground_states} (see also ref.~\onlinecite{Likos_comment_on_Suto}) has shown rigorously that integrable bounded potentials whose Fourier transform is non-negative and vanishes above a wave number $K_0$ have an infinite number of continuously degenerate ground states above a well-defined threshold density. This means that, for this class of pair potentials, the potential energy is completely flat along certain directions in configuration space. Although the Hertz potential does not satisfy the S\"ut\H{o} criteria (its Fourier transform does not vanish above any $K_0$), its Fourier transform does decay $\propto k^{-4}$ for wave vectors larger than $\sim 10\sigma^{-1}$ (see inset in Fig.~\ref{fig:phase_diagram}). We speculate that, as a result, certain collective motions in the dense phase cost very little potential energy. It is the softness of the pair potential that causes melting into a rather peculiar liquid for which particle motions should be strongly correlated.

\begin{figure}
    \includegraphics[scale=.475]{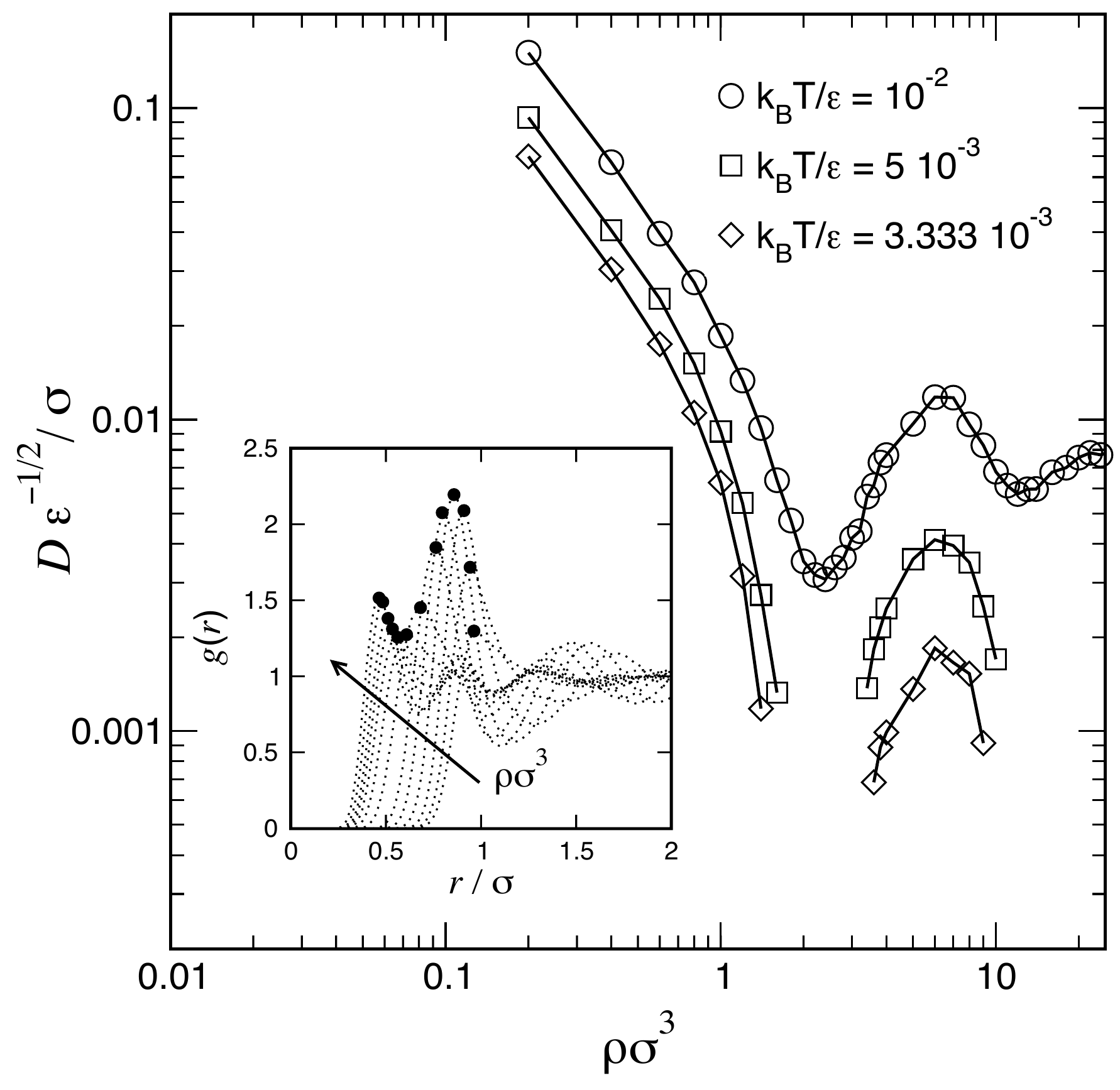}
    \caption{\label{fig:diffusivity} Diffusion coefficient of the fluid phase obtained by molecular dynamics as a function of density at three temperature values. Breaks in the lines joining the simulation data indicate that points in between were identified as crystalline states. The inset shows the radial distribution function $g(r)$ for the isotherm at $k_BT/\epsilon=0.01$ and the following values of $\rho\sigma^3$: 0.5, 1, 1.5, 2, 2.5, 3, 4, 5, 6, 7, 8, 10, 12. The highest peak in each $g(r)$ is marked with a black dot. In the set of black dots, the local maximum and minimum occur approximately at the densities at which the diffusion coefficient has a minimum ($\rho\sigma^3$=2.40) and a maximum ($\rho\sigma^3$=6.35), respectively.}
\end{figure}

In Fig.~\ref{fig:diffusivity} we show other aspects of the peculiarity of the dynamics of the fluid phase. The diffusion coefficient $D$, which decreases monotonically with increasing density in simple liquids, shows a clear non-monotonic behavior in the Hertzian fluid. Moreover, the minimum in $D$ at $k_BT/\epsilon=10^{-2}$ coincides with the maximum freezing point at $\rho\sigma^3=2.40$, and a maximum in the first peak of the $g(r)$ (see inset in Fig.~\ref{fig:diffusivity}). Also the maxima in $D$ for the three temperatures shown in Fig.~\ref{fig:diffusivity} are located at the same density as the maximum freezing point at $\rho\sigma^3=6.35$ and a minimum in the first peak of $g(r)$. These multiple structural and dynamical anomalies, and their interrelation, has been seen before in bounded potentials,~\cite{GCM_anomalies, connection_dynamics_phase_diagram} in liquids with water-like anomalies~\cite{structural_order_anomalies} and colloids with short-ranged attractions.~\cite{colloids_attraction_anomalies} This suggests that crystallization and fluid dynamics are inextricably connected, and that this connection is not independent of the underlying crystal structure. For instance, the presence of the H and SC crystals does not imprint peaks in the diffusion coefficient as the BCT does. We think that this is related to the fact that single-particle mobility is affected by differences in packing efficiency: the intrinsic volume of the BCT primitive cell is much smaller than that of the BCC, H and SC structures. Besides, even though particle motions in the low-temperature liquid phase are strongly correlated, the liquid does not appear to be particularly glassy. In fact, within the range of densities studied, we observe spontaneous crystallization rather than structural arrest upon cooling even at temperatures below $k_BT<10^{-4}$. Further study of the dynamics of the high-density, supercooled liquid would therefore be most interesting.

Another interesting feature of the system at high densities and low temperatures is its freezing behavior. We have used free-energy calculations to trace freezing curves up to a density of $\rho\sigma^3\approx 7$. Even beyond that density, there appears to be a succession of other freezing transitions. Although we have not traced the accurate phase boundaries completely for $\rho\sigma^3>7$, we have calculated free energies at $\rho\sigma^3=9$ and $k_BT/\epsilon=10^{-3}$, and found the stable structure to be the BCT with $c/a \approx 1$. Additionally, we have found that crystallization occurs at low temperatures and at densities up to $\rho\sigma^3=12$. In other words: as far as we can tell, the low-temperature phase is always crystalline, but the structure of the stable crystalline phase changes as the density is increased. There is also evidence~\cite{Hertz_ground_states} at $T=0$ supporting this observation. In fact, the ground-state behavior is interesting in the context of a result reported by Torquato and Stillinger:~\cite{duality_relations} on the basis of duality relations between a soft potential and its Fourier transform, these authors found a one-dimensional system, the overlap potential, that exhibits an infinite number of phase transitions between periodic ground states over the entire density range. One can then wonder whether the Hertz potential also shows an unbounded number of phase transitions between periodic structures. Interestingly, the Hertz potential and the overlap potential in 1D and 3D, $V_{1D}(r)=\epsilon(1-r/\sigma)$ and $V_{3D}(r)=\epsilon(1-(3/2)(r/\sigma)+(1/2)(r/\sigma)^3)$ for $r<\sigma$, studied in ref.~\onlinecite{duality_relations}, belong to the same class of non-negative, bounded functions, and both have oscillatory, decaying Fourier transforms (see ref.~\onlinecite{url_Wolfram_demo} and inset in Fig.~\ref{fig:phase_diagram}).

In summary, we have shown that the Hertz potential gives rise to a phase diagram with multiple re-entrant melting transitions and a succession of Bravais crystals spanning a wide range of densities. In addition, the diffusion coefficient of the re-entrant fluid shows unusual non-monotonic curves with increasing density. This rich behavior together with the simplicity of the potential's functional form makes the Hertz model an attractive one for the study of kinetic phenomena in soft-core systems, like martensitic transformations and supercooled dynamics.

\begin{acknowledgments}

J.C.P. thanks Gerhard Kahl and Bianca Mladek for their help during a stay at TU Wien. Financial support from the EU contract MRTN-CT-2003-504712 and the short-term scientific mission COST-STSM-P13-02014 are acknowledged. A.C. acknowledges financial support from Columbia University. The work of the FOM Institute, where most of this work has been carried out, is part of the research program of FOM and is made possible by financial support from the Netherlands organization for Scientific Research (NWO).

\end{acknowledgments}


\begin{thebibliography}{35}
\bibitem[{not({\natexlab{a}})}]{note2}
\bibinfo{note}{For spheres of identical size,
  $\epsilon=\frac{2}{15}\frac{E\sigma^3}{(1-\nu^2)}$, where $E$ and $\nu$ are
  respectively the Young modulus and the Poisson ratio of the material the
  spheres are made of, and $r$ is their center-to-center distance. This
  functional form for the elastic energy is based on linear elasticity theory
  and geometric considerations (see ref.~\onlinecite{Landau&Lifshitz}).}
\bibitem[{not({\natexlab{b}})}]{note1}
\bibinfo{note}{The primitive unit cell, which is the smallest possible unit
  cell, does not necessarily coincide with the conventional unit cell. For
  example, in the FCC lattice, all angles in the conventional unit cell are
  square, and $V/a^3=1$; its primitive cell, however, has angles of 45$^\circ$
  and 60$^\circ$, and $V/a^3=1/4$.}
\end{thebibliography}
\end{document}